\documentclass[12pt]{article}
\usepackage{amsmath,amssymb}
\newcommand{\beq}{\begin{equation}}
\newcommand{\eeq}{\end{equation}}
\newcommand{\ov}{\overline}
\newcommand{\pa}{\partial}
\newcommand{\al}{\alpha}

\newcommand{\lan}{\langle}
\newcommand{\ran}{\rangle}

\renewcommand{\thefootnote}{\fnsymbol{footnote}}

\begin{document}

\title{
\begin{flushright}
\ \\*[-80pt] 
\begin{minipage}{0.2\linewidth}
\normalsize
KUNS-2130 \\*[50pt]
\end{minipage}
\end{flushright}
{\Large \bf 
Soft supersymmetry breaking terms \\ from 
$D_4 \times Z_2$ lepton flavor symmetry 
\\*[20pt]}}

\author{
\centerline{
Hajime~Ishimori$^{1,}$\footnote{E-mail address: ishimori@muse.sc.niigata-u.ac.jp},   
~Tatsuo~Kobayashi$^{2,}$\footnote{E-mail address: kobayash@gauge.scphys.kyoto-u.ac.jp}, 
~Hiroshi~Ohki$^{3,}$\footnote{E-mail address: ohki@scphys.kyoto-u.ac.jp},} \\ 
\centerline{
~Yuji~Omura$^{3,}$\footnote{E-mail address: omura@scphys.kyoto-u.ac.jp},  
~Ryo~Takahashi$^{1,}$\footnote{E-mail address: takahasi@muse.sc.niigata-u.ac.jp} \ and \
Morimitsu~Tanimoto$^{4,}$\footnote{E-mail address: tanimoto@muse.sc.niigata-u.ac.jp} }
\\*[20pt]
\centerline{
\begin{minipage}{\linewidth}
\begin{center}
$^1${\it \normalsize
Graduate~School~of~Science~and~Technology,~Niigata~University, \\ 
Niigata~950-2181,~Japan } \\
$^2${\it \normalsize 
Department of Physics, Kyoto University, 
Kyoto 606-8502, Japan} \\
$^3${\it \normalsize 
Department of Physics, Kyoto University, 
Kyoto 606-8501, Japan} \\
$^4${\it \normalsize
Department of Physics, Niigata University,~Niigata 950-2181, Japan } 
\end{center}
\end{minipage}}
\\*[50pt]}

\date{
\centerline{\small \bf Abstract}
\begin{minipage}{0.9\linewidth}
\medskip 
\medskip 
\small
We study the supersymmetric model with $D_4 \times Z_2$ 
lepton flavor symmetry. 
We evaluate soft supersymmetry breaking terms, i.e. 
soft slepton masses and A-terms, which are predicted 
in the $D_4$ flavor model.
We consider constraints due to experiments of 
flavor changing neutral current processes.     
\end{minipage}
}

\begin{titlepage}
\maketitle
\thispagestyle{empty}
\end{titlepage}


\renewcommand{\thefootnote}{\arabic{footnote}}
\setcounter{footnote}{0}

\section{Introduction}

Understanding the origin of fermion flavor structure, i.e. 
quark/lepton masses and mixing angles, is 
one of important issues in particle physics.
Indeed, various types of mechanisms have been proposed 
to realize realistic mass matrices of quarks and leptons.
Hereafter, we refer to such mechanisms as flavor mechanisms.
Non-Abelian discrete flavor symmetries are 
interesting proposals and several types of models with 
non-Abelian discrete flavor symmetries have been constructed \cite{Discrete}.

Origins of such non-Abelian discrete flavor symmetries 
are not clear in 4D field theory, but those may be 
originated from geometrical structures of extra 
dimensional field theories and superstring theories 
on 6D compact spaces.
In Ref.~\cite{Kobayashi:2004ya,Kobayashi:2006wq,Ko:2007dz} it 
has been shown that certain flavor structures can be 
derived from heterotic string models on orbifolds.
One of typical non-Abelian discrete flavor symmetries, 
which can appear from heterotic orbifold models, is 
$D_4$ symmetry and it can appear from factorizable 
orbifolds including the $Z_2$ orbifold.
Indeed, several semi-realistic string models 
with the $D_4$ flavor symmetry have been constructed 
in Ref.~\cite{Kobayashi:2004ya,Kobayashi:2004ud}, where 
three families consist of $D_4$ singlets and doublets.
Thus, it is important to study several phenomenological 
aspects of $D_4$ flavor models.

Grimus and Lavoura proposed first the $D_4$ flavor model 
for the lepton sector \cite{Grimus} and subsequently 
other several models were studied \cite{D4,D4-2}.
Recently, the authors proposed a new $D_4$ lepton flavor 
model \cite{Ishimori:2008gp}, 
which has only a single electroweak Higgs field, 
while the Grimus-Lavoura model as well as other models 
has three electroweak Higgs fields.

Supersymmetric extension of the standard model (SM) is 
one of interesting candidates for new physics beyond a TeV scale.
Supersymmetry (SUSY) can stabilize the Higgs mass against 
radiative corrections due to heavy modes around high energy scales, 
e.g. the right-handed Majorana neutrino mass scale, the GUT scale, 
the Planck scale, etc.
Supersymmetric standard models also have a good candidate 
for dark matter.
Furthermore, in the minimal supersymmetric standard model (MSSM) 
with a pair of up and down Higgs supermultiplets, 
three gauge couplings are unified at the GUT scale 
$M_X \simeq 2 \times 10^{16}$ GeV in a good accuracy.

In supersymmetric models, quarks and leptons have 
their superpartners, i.e. squarks and sleptons.
If we have a  flavor mechanism to lead to 
realistic masses and mixing angles of quarks and leptons, 
such a flavor mechanism would affect mass matrices 
of squarks and sleptons.
Furthermore, if a specific pattern of sfermion 
masses is derived by a certain flavor mechanism, 
that would become the prediction of a certain flavor 
mechanism, which could be tested by measuring 
squark/slepton masses in future experiments.
Thus, it is important to study patterns of sfermion 
masses, which are obtained by flavor mechanisms 
leading to realistic fermion masses and mixing angles.
Although squarks and sleptons 
have not been detected yet, their mass matrices 
are constrained severely by experiments of 
flavor changing neutral current (FCNC) 
processes~\cite{FCNCbound}.\footnote{See also 
e.g  Ref.~\cite{Chankowski:2005jh} and references therein.}
Off-diagonal elements of sfermion mass matrices 
must be suppressed in the super-CKM basis, where 
fermion masses are diagonalized, when 
sfermion masses are of order of the weak scale.
At any rate, it is important to study which patterns of 
sfermion mass matrices, i.e. soft scalar masses and 
A-terms,  are derived from each flavor mechanism.

In this paper, we study supersymmetric extension of 
the $D_4$ model of Ref.~\cite{Ishimori:2008gp} 
as well as the Grimus-Lavoura model.
The $D_4$ model of Ref.~\cite{Ishimori:2008gp} has a single 
electroweak Higgs field, 
while other models have more than one electroweak Higgs fields.
This difference is important in supersymmetric extensions.
The former becomes the MSSM with a pair of 
up and down Higgs fields at low energy, but 
the latter would have more pairs of Higgs fields.
The latter may violate the gauge coupling unification 
unless we introduce extra colored supermultiplets.
That is the reason why we study mainly supersymmetric extension of 
the $D_4$ model of Ref.~\cite{Ishimori:2008gp}.
We evaluate soft SUSY breaking terms of 
sleptons, which are derived in supersymmetric 
extension of the $D_4$ model~\cite{Ishimori:2008gp}  
as well as the Grimus-Lavoura model.
We compute soft scalar masses and A-terms for the lepton sector 
in these supersymmetric models within the framework of 
the gravity mediation of SUSY breaking.
We examine constraints due to FCNC experiments.\footnote{
See Ref.~\cite{Ko:2007dz} for a similar analysis on the 
quark sector.}

The paper is organized as follows.
In Section 2, we supersymmetrize the $D_4$ model of~\cite{Ishimori:2008gp} and 
show values of parameters consistent with
neutrino oscillation experiments.
In Section 3, we evaluate soft SUSY breaking terms of 
sleptons, i.e. soft scalar mass matrices and A-terms.
We examine FCNC constraints on those SUSY breaking terms 
as mass insertion parameters.
Section 4 is devoted to conclusion and discussion.
In Appendix, we also discuss supersymmetric extension of the 
Grimus-Lavoura model.
We evaluate soft SUSY breaking terms in the supersymmetric 
Grimus-Lavoura model and show that it is almost the same 
as the results obtained in Section 3.

\section{Supersymmetric model with $D_4$ flavor symmetry}

In this section, we study  our supersymmetric $D_4$ model, 
based on \cite{Ishimori:2008gp}.
The $D_4$ symmetry has five irreducible representations, i.e. 
one doublet ${\bf 2}$ and four singlets ${\bf 1}_{++}$, 
${\bf  1}_{+-}$,  ${\bf 1}_{-+}$ and ${\bf 1}_{--}$, 
where ${\bf 1}_{++}$ is the trivial singlet and 
the others are non-trivial singlets.
Their tensor products are obtained as 
\beq
{\bf 2} \otimes {\bf 2} = {\bf 1}_{++} \oplus {\bf 1}_{+-} \oplus
{\bf 1}_{-+} \oplus {\bf 1}_{--},  \qquad 
{\bf 1}_{ab} \otimes
{\bf 1}_{cd} = {\bf 1}_{ef} ,
\eeq 
where $a,b,c,d=\pm$ and $e=ac$ and $f=bd$.

The three generations of left-handed, right-handed charged lepton and
right-handed neutrino chiral superfields are denoted by 
$L_I,~R_I,~N_I,~(I=e,~\mu,~\tau)$, respectively, and are 
assigned to $D_4$ trivial singlets ${\bf 1}_{++}$ and doublets 
${\bf  2}$. 
We also introduce a $D_{4}$ doublet, $(\chi_1,~\chi_2)$, a $D_4$
non-trivial singlet, $\chi_{-+}$, and a $D_4$ trivial singlet,
$\chi$. 
They are all SM-gauge singlet chiral superfields.


Moreover, we introduce additional $Z_2$ symmetry 
and we assume that the superfields
$R_e,(R_{\mu},~R_{\tau}),~\chi,~\chi_{-+}$ have $Z_2$-odd charges.
The others are $Z_2$-even.
Higgs chiral superfields for both up and down sectors, $H^{u,d}$, 
are assumed to be $D_4$ trivial singlets and $Z_2$-even. 
These assignments are shown in Table 1.
Hereafter, we follow the conventional notation that 
superfields and their scalar components are denoted 
by the same letters.
\begin{table}[htb]
\begin{center}
\begin{tabular}{|c|cccccc||c||ccc|}
\hline
              &$L_e$   & $L_I$ & $R_e$  & $R_I$ & $N_e$ & $N_I$ &  $H^{u,d}$&$\chi$ &$\chi_{-+}$ &$(\chi_1,\chi_2)$   \\ \hline
$D_4$      &${\bf 1}_{++}$       & ${\bf 2}$                    & ${\bf
  1}_{++}$      &                ${\bf 2}$ &       ${\bf 1}_{++}$ & ${\bf 2}$
& ${\bf 1}_{++}$         &${\bf 1}_{++}$&${\bf 1}_{-+}$& ${\bf 2}$      \\
$Z_2$ &   +     &          +              &  $-$ &      $-$
&     +         &            +                & + &$-$&$-$ & + \\ \hline
\end{tabular}
\end{center}
\caption{$D_4$ and $Z_2$ charges. $I$ corresponds to $I=\mu$ and $\tau$.
\label{charge}}
\end{table}

The $D_4 \times Z_2$ invariant superpotential of the charged leptons 
is given as
\beq
W^{(4)}_L=\frac{y_e}{M_p} L_e R_e H^d \chi +\frac{y_{\mu}}{M_p} (L_{\mu}R_\mu +L_{\tau}R_\tau)H^d \chi +\frac{y'_{\mu}}{M_p} (L_{\mu}R_\mu-L_{\tau}R_\tau) H^d \chi_{-+}, 
\label{W-l}
\eeq
up to 5-point couplings, 
where $M_p$ is the Planck scale, $M_p=2.4 \times 10^{18}$ GeV. 
We assume that scalar components of $\chi_1, \chi_2, \chi$ and
$\chi_{-+}$ develop their vacuum expectation values (VEVs) 
and the $D_4 \times Z_2$ symmetry is broken at a high energy scale. 
Then, the above superpotential (\ref{W-l})  leads to a diagonal charged 
lepton mass matrix after the electroweak symmetry breaking 
by Higgs VEVs, $v^d = \langle H^d \rangle$ and $v^u = \langle H^u \rangle$. 
On the other hand, higher-order operators can be also included 
in the superpotential,
\beq
W^{(5)}_L=\frac{y_{\mu e}}{M_p^2} L_e (R_{\mu}\chi_1 +R_{\tau} \chi_2 ) 
\chi H^d + \cdots,
\eeq
and these generate off-diagonal elements of the charged lepton mass
matrix. 
As a result, after $D_4\times Z_2$ symmetry breaking, 
the effective Yukawa matrix $(\tilde y_\ell)_{IJ}$ among the charged leptons 
and the Higgs field $H^d$ is found to be  
\beq
\begin{split}
(\tilde y_\ell)_{IJ}=& 
\left(
  \begin{array}{ccc}
y_e\alpha_a  &  y_{e\mu}\alpha\alpha_a-y'_{e\mu}\alpha\alpha_b &  y_{e\mu}\alpha\alpha_a+y'_{e\mu}\alpha\alpha_a \\ 
y_{\mu e}\alpha\alpha_a-y'_{\mu e}\alpha\alpha_b    & y_\mu\alpha_a -y'_{\mu}\alpha_b  & y_{\mu\tau}\alpha_a\alpha^2+y'_{\mu\tau}\alpha_b\alpha^2     \\
y_{\mu e}\alpha\alpha_a+y'_{\mu e}\alpha\alpha_b    & y_{\mu\tau}\alpha_a\alpha^2-y'_{\mu\tau}\alpha_b\alpha^2 & y_\mu\alpha_a+y'_{\mu}\alpha_b    \\  
\end{array} \right),
\end{split}
\label{eq:clepton}
\eeq
where $\alpha=\lan \chi_i \ran /M_p$, $\alpha_a=\lan \chi \ran /M_p$,
$\alpha_b= \lan \chi_{-+}\ran /M_p$.
Here, it has been assumed that 
$\lan \chi_1 \ran = \lan \chi_2 \ran$~\cite{Ishimori:2008gp}.
The mass matrix of the charged leptons is obtained as 
$M_\ell =  v^d \ \tilde y_\ell$.
If $\al$ is allowed sufficiently small compared with $m_e/m_{\tau}$, 
charged lepton masses are estimated by the diagonal elements of 
(\ref{eq:clepton}) as follows,
\beq
m_e \sim  y_e \alpha_a v^d, \qquad m_{\mu} \sim 
( y_\mu\alpha_a -y'_{\mu}\alpha_b ) v^d,~\qquad  
m_{\tau} \sim
(y_\mu\alpha_a + y'_{\mu}\alpha_b)v^d.
\eeq
We need the fine-tuning of parameters, 
$y_\mu$, $y'_\mu$, $\alpha_a$ and $\alpha_b$, such that we realize 
the mass ratio $m_\mu/m_\tau$ as 
$( y_\mu\alpha_a -y'_{\mu}\alpha_b )/(y_\mu\alpha_a +
y'_{\mu}\alpha_b) = {\cal O}(m_\mu/m_\tau)$.
That is similar to Ref.~\cite{Grimus}. 
Also, the coupling $y_e$ must be suppressed to lead to 
$y_e \alpha_a/(y_\mu\alpha_a +
y'_{\mu}\alpha_b) = {\cal O}(m_e/m_\tau)$.
Off-diagonal elements of the diagonalizing matrix of 
(\ref{eq:clepton}) are determined by the parameter $\alpha$.
For example, the (1,2) element of the diagonalizing matrix 
is estimated as 
\beq
\theta^\ell_{12} \sim \frac{( y_{e\mu}\alpha_a -y'_{e\mu}\alpha_b ) \alpha 
v^d}{m_\mu} , 
\label{theta-12-1}
\eeq
and it becomes 
\beq
\theta^\ell_{12} \sim \al m_{\tau}/m_{\mu},
\label{theta-12-2}
\eeq
when $y_{e\mu}, y'_{e\mu} \sim 1$.
At any rate, when $\alpha \leq {\cal O}(10^{-2}) $, 
the mixing angle $\theta^\ell_{12}$ is small.\footnote{
Small deviations from the diagonal form are important 
in a certain case (see e.g. \cite{Hochmuth:2007wq}).}
Such a small value is not important for the neutrino oscillation 
in our model at the current level of experiments, 
but important for soft SUSY breaking terms as we will discuss.

Now we consider the neutrino sector.
The $D_4 \times Z_2$-invariant superpotential is given as
\beq
\begin{split}
W_N^{(3)}=&~y_1N_e L_e H^u + y_2(N_\mu L_\mu + N_\tau L_\tau )H^u \\
  &+y_a N_e(N_{\mu}\chi_1+N_{\tau}\chi_2) 
  +M_1 N_eN_e+M_2(N_{\mu}N_{\mu}+N_{\tau}N_{\tau}),
\end{split}  
\eeq
up to 4-point couplings.
The higher-order operators should be also considered
\beq
\label{matrix}
W^{(4)}_N=\frac{y_{21}}{M_p} N_e (L_{\mu}\chi_1 +L_{\tau} \chi_2 ) H^u+
\frac{y_{12}}{M_p} L_e (N_{\mu}\chi_1 +N_{\tau} \chi_2 )  H^u + \cdots.  
\eeq
With the above superpotential, the Majorana mass matrix, $M_R$, 
and the Dirac mass matrix, $M_D$, of the right-handed neutrinos are 
written down as  
\beq
\begin{split}
M_R
=& 
\left(
  \begin{array}{ccc}
M_1   &  y_a M_p \alpha &  y_a M_p\alpha  \\ 
            y_a   M_p \alpha    & M_2  & y_{b}M_p \alpha^2    \\
            y_a   M_p \alpha   & y_{b}M_p \alpha^2 & M_2     \\  
\end{array} \right), \\
M_D
=& v^u
\left(
  \begin{array}{ccc}
                  y_{1}    & y_{12}\alpha  & y_{12}\alpha  \\ 
                    y_{21}\alpha    & y_2  & y_{23}\alpha^2    \\
                    y_{21}\alpha     & y_{32}\alpha^2  & y_2   \\
  \end{array} \right),
\end{split}
\label{MR-MD}
\eeq
after the $D_4 \times Z_2$ symmetry and the electroweak symmetry 
are broken.
The light neutrino mass matrix is obtained as,
\beq
M_{\nu}=M_D^TM_R^{-1}M_D=
\begin{pmatrix}x & y & y \\ y & z & w \\ y & w & z  \end{pmatrix},  
\eeq
\noindent
 where $x$, $y$, $z$ and $w$ are given in terms of parameters of
 Eq.(\ref{MR-MD}).
This form of $M_{\nu}$ can realize the realistic neutrino mixing 
without fine-tuning~\cite{Grimus,D4,D4-2,Ishimori:2008gp}. 
The light neutrino masses are written as\footnote{
These values are obtained at the energy scale $M_R$ 
and radiative corrections are, in general, important to 
evaluate values at low energy.
However, radiative corrections are negligible in this $D_4$ 
flavor model~\cite{D4-2}.}
\beq
M_{\nu}  =  V
  \begin{pmatrix}
m_1   &  0 &  0 \\ 
                     0    & m_2  & 0     \\
                     0   & 0 & m_3    \\  
\end{pmatrix} 
  V^T, 
\eeq
\beq
V = \begin{pmatrix} 
\cos \theta_{12}    &  \sin \theta_{12} &  0 \\ 
                     -\sin \theta_{12} /\sqrt{2}    & \cos \theta_{12} /\sqrt{2}  & 1/\sqrt{2}     \\
                     -\sin \theta_{12}/\sqrt{2}   & \cos \theta_{12} /\sqrt{2} & -1/\sqrt{2}    \\  
\end{pmatrix} ,
\label{mixing-V}
\eeq
\begin{eqnarray}
m_1&=&
\frac12\cdot
\frac{y_a^2r^2+y_2^2r+\sqrt2 y_1y_2y_akr/(\cos \theta_{12} \sin \theta_{12})}
{r^2-2y_a^2k^2r}\cdot
\frac{(v^u)^2}{M_1} ,
\nonumber\\
m_2&=&
\frac12\cdot
\frac{y_a^2r^2+y_2^2r-\sqrt2 y_1y_2y_akr/(\cos \theta_{12} \sin \theta_{12})}
{r^2-2y_a^2k^2r}\cdot
\frac{(v^u)^2}{M_1} ,
\nonumber\\
m_3&=&
\frac{y_2^2(r-2 y_a^2k^2)}
{r^2-2y_a^2k^2r}\cdot
\frac{(v^u)^2}{M_1},
\label{mass2}
\end{eqnarray}
where we define as
$r\equiv \frac{M_2}{M_1}$ and $k\equiv \frac{\alpha M_p}{M_1}$, 
and neglect higher order terms of $\alpha$ such 
as $\alpha^2M_p$ and $\alpha^2M_1$ appeared in $M_R^{-1}$. 
The justification of this approximation and 
derivation of Eq. (\ref{mass2}) are given in Ref.~\cite{Ishimori:2008gp}. 
The mixing angle $\theta_{12}$ is written as 
\beq
\cot2\theta_{12} =
\frac{y_1^2r-y_2^2}
{2\sqrt2 y_1y_2y_ak}\ .
\label{angle}
\eeq
When $y_1,y_2$ and $y_a$ are of ${\cal O}(1)$, the above mixing angle 
$\theta_{12}$ is of ${\cal O}(1)$ and the effect due to
$\theta^\ell_{12}$ is negligible in the neutrino oscillation.
Thus,  the atmospheric neutrino mixing angle 
is maximal and the Chooz mixing angle is vanishing, while the solar neutrino 
mixing is of ${\cal O}(1)$. 
This  also holds if the mass matrices
were complex, i.e. if the CP violating case would be studied.

Now, let us consider the realization of experimental values.
We use the best fit values of mass squared differences and solar
mixing angle as \cite{neutrino}
\begin{eqnarray}
 & & \Delta m_{\rm atm}^2=2.4\times10^{-3}{\rm~~eV}^2, \qquad 
\Delta m_{\rm sol}^2=7.6\times10^{-5}{\rm~~eV}^2, \nonumber \\
 & &  \sin^2\theta_{12}=0.32 .
\label{best-fit}
\end{eqnarray}
First we consider the simple case with  $y_1=y_2=y_a=1$.
The above experimental values (\ref{best-fit}) are obtained by 
taking\footnote{
If $M_1,~M_2,~M_p \al$ satisfy the relation $M_1+ M_p \al=M_2$, 
$\cot 2\theta_{12}=1/(2\sqrt{2})$ is realized 
at this approximation level
and the mixing matrix $V$ 
in Eq.(\ref{mixing-V}) is the
so-called tri-bimaximal matrix~\cite{tri-bimixing}. }
\beq
M_1 = 4.9\times10^{15}{\rm GeV} , \qquad M_2 = 6.2\times10^{14}{\rm GeV} ,
\qquad |\al| = 1.6 \times10^{-3}, 
\label{eq:value}
\eeq
for $v^u \simeq 174$~GeV.\footnote{The numerical result  has been
given in the non-SUSY case \cite{Ishimori:2008gp}. }
In this parametrization, the parameter $\al$ is sufficiently small, 
so that the charged lepton mass matrix is diagonal as we
assumed in the beginning of this section. 
When we vary $y_1,~y_2$ and $y_a$ around $y_1,~y_2,~y_a={\cal O}(1)$, 
the above experimental (\ref{best-fit}) values are realized 
for~\cite{Ishimori:2008gp}
\begin{eqnarray}
 & & M_2 \sim 0.2 \times \left(
\frac{y_2}{y_1}\right)^2 M_1 , \qquad 
 M_1 
\sim 3\times 10^{15} \times y_1^2 {\rm~~GeV} , \nonumber \\
 & &  \alpha \sim  0.001 \times \frac{y_1 y_2}{y_a} .
\label{eq:value2}
\end{eqnarray}    
Thus, it is found that the value of $\al$ is predicted 
around ${\cal O}(10^{-4})-{\cal O}(10^{-2})$ as long as 
couplings, $y_1, y_2, y_a$  are of ${\cal O}(1)$.

\section{Soft SUSY breaking terms in supersymmetric $D_4$ model}

We consider soft SUSY breaking terms of slepton mass matrices 
within the framework of supergravity theory, 
i.e. the gravity mediation.\footnote{SUSY breaking may be realized 
as the gauge mediation or anomaly mediation.
These mediation mechanisms would lead to the flavor-blind soft SUSY 
breaking terms.}
In our model, the $D_4$ symmetry restricts not only the fermion mass
matrices but also the scalar matrices. 
Now we assume chiral superfields $\Phi_k$ to cause SUSY breaking 
by their non-vanishing F-components. The F-components are given by 
\beq
F^{\Phi_k}= - e^{ \frac{K}{2M_p^2} } K^{ \Phi_k \ov{I} } \left(
  \pa_{\ov{I}} \ov{W} + \frac{K_{\ov{I} }} {M_p^2} \ov{W} \right) ,
\label{eq:F-component}
\eeq
where $K$ denotes the K\"ahler potential, $K_{\ov{I}J}$ denotes 
second derivatives by fields, i.e. $K_{\ov{I}J}={\pa}_{\ov{I}} \pa_J K$
and $K^{\ov{I}J}$ is its inverse. 
In general, the fields $\Phi_k$ in our notation include 
$D_4 \times Z_2$-singlet moduli fields $Z$ and $\chi,~\chi_{-+},~\chi_i$.
Furthermore VEVs of $F_{\Phi_k}/\Phi_k$  are estimated as 
$\lan F_{\Phi_k}/ \Phi_k \ran = {\cal O}(m_{3/2})$, where
$m_{3/2}$ denotes the gravitino mass, which is obtained as 
$m_{3/2}= \lan e^{K/2M_p^2}W/M_p^2 \ran$.

First let us discuss the scalar mass matrices given by using the second-order K\"{a}hler potential of left-handed and right-handed leptons, 
\beq
\begin{split}
K^{(2)}=& a_e(\Phi_k ) L_{e}^{\dagger} L_e +a_{\mu}(\Phi_k)(L_{\mu}^{\dagger}L_{\mu}+L_{\tau}^{\dagger}L_{\tau}) \\
        &+b_e(\Phi_k) R_e^{\dagger} R_e +b_{\mu}(\Phi_k)(R_{\mu}^{\dagger} R_{\mu}+ R_{\tau}^{\dagger} R_{\tau}),  
\end{split}
\eeq
where $a_{e,\mu}(\Phi_k),~b_{e,\mu}(\Phi_k)$ are 
$D_4 \times Z_2$-invariant generic functions. 
Then the soft SUSY breaking scalar masses are given 
by \cite{Kaplunovsky:1993rd}
\beq
m^2_{\ov{I}J} K_{\ov{I}J} = m_{3/2}^2 K_{\ov{I}J} - |F^{\Phi_k}|^2 \pa_{\Phi_k}  
\pa_{  \ov{\Phi_k} }  K_{\ov{I}J} + 
|F^{\Phi_k}|^2 \pa_{\ov{\Phi_k}}  K_{\ov{I}L} \pa_{\Phi_k}  
K_{\ov{M} J} K^{L \ov{M}},
\label{eq:scalar}
\eeq
for the scalar fields with the K\"ahler metric $K_{\ov{I}J}$, 
where we have assumed the vanishing vacuum energy.
Evaluating $\lan F^{\Phi_k}/  \Phi_k \ran $ for all fields as
$\lan F^{\Phi_k}/  \Phi_k \ran ={\cal O}(m_{3/2})$, 
the slepton mass matrices can be found to be  
\beq
{m}_{L}^2=
\begin{pmatrix} 
m_{L1}^2 & 0 & 0 \\ 
0 & m_{L2}^2    & 0 \\ 
0 & 0 &  m_{L2}^2 
\end{pmatrix},~ \qquad 
{m}_{R}^2=
\begin{pmatrix} 
m_{R1}^2 & 0 & 0 \\ 
0 & m_{R2}^2    & 0 \\ 
0 & 0 &  m_{R2}^2 
\end{pmatrix},
\label{scalar-1}
\eeq
where $m_{Li},~m_{Ri}={\cal O}(m_{3/2})$. 
These contributions give the leading-order of the slepton 
mass matrices.
Since the second and third families are $D_4$ doublets,  
these forms (\ref{scalar-1})
can be easily expected from the $D_4$ flavor symmetry 
and it is the prediction of our model that the second and third families of 
sleptons have degenerate masses at this level.\footnote{
When three families consist of a singlet and a doublet under 
a non-Abelian discrete symmetry, a similar structure would appear, 
that is, two of three families of scalar masses corresponding to 
the doublet would be degenerate at a certain level.
See e.g. Ref.~\cite{Kobayashi:2003fh}.
However, the exactly same structure as our model has 
not been predicted in other models.
}

Eq.~(\ref{scalar-1}) is written in the $D_4$ flavor basis.
The super-CKM basis is convenient to examine constraints due to 
FCNC.
For example, the (1,2) elements are obtained in the super-CKM basis as 
$(m_{L(R)}^2)_{12}^{(SCKM)} \sim \theta^\ell_{12}(m_{L(R)1}^2-m_{L(R)2}^2)$.
The mass insertion parameters are defined as 
\beq
(\delta^l_{LL(RR)})_{ij} \equiv \frac{(m_{L(R)}^2)_{ij}^{(SCKM)}}{m_{SUSY}^2} ,
\eeq
where $m_{SUSY}$ denotes the average mass of sleptons.
The (1,2) elements are constrained severely by 
the $\mu \rightarrow e \gamma$ experiments as  
$(\delta^l_{LL(RR)})_{12} \leq {\cal O}(10^{-3})$ when 
$m_{SUSY} ={\cal O}(100)$ GeV, while the others have no strong 
constraints.
That requires 
\beq
\theta^\ell_{12} \leq {\cal O}(10^{-3}).
\label{constraint-12}
\eeq
For $y_1=y_2=y_a=1$,
we obtain $\alpha ={\cal O}( 10^{-3})$ and 
$\theta_{12}^\ell= {\cal O}(10^{-2})$, and  
such a parameter region will be ruled out. 
The value of 
$\alpha$ is obtained 
as $\alpha = 0.001 \times y_1y_2/y_a$ 
for $y_1, y_2, y_a \sim 1$ and
we estimate $\theta_{12}^\ell=\alpha y_{e\mu}$  
 for  $y_{e \mu } \sim y'_{e \mu }$ 
as discussed in Section 2.
Possibly, we can get $\theta_{12}^\ell= {\cal O}(10^{-3})$ in a certain 
parameter region, e.g. $y_1=y_2=y_{e \mu } =y'_{e \mu }=1/y_a=0.5$.
Hence, our model is marginal for FCNC constraints, 
that is, a certain parameter region is ruled out already but 
the remaining parameter region is still wide.
Of course, if the couplings $y_{e \mu }, y'_{e \mu }$,
which are irrelevant to realization for the neutrino oscillation 
experiments,  are suppressed 
as $y_{e \mu }, y'_{e \mu } \ll {\cal O}(1)$, 
a quite wider region for $y_1, y_2, y_a$ is allowed.

In the above estimation we restricted the K\"{a}hler potential to
$K^{(2)}$, but we can write more general terms in the K\"{a}hler potential 
after $D_4 \times Z_2$-symmetry breaking.
The VEVs of $\chi_i$ are large compared with VEVs of the other fields $\chi$
and $\chi_{-+}$ as the result of Section 2.
Thus, corrections including $\chi_i$ are more important 
in correction terms of the K\"{a}hler potential.
Therefore, we include such terms in the K\"{a}hler potential.
For example, the following terms  
\begin{equation}
\Delta K_L^{(e)}=\frac{\beta_{L1}(Z)}{M_p}
 L_e^{\dagger}(L_{\mu}\chi_1^{\dagger}+L_{\tau}\chi_2^{\dagger})
+ \frac{\beta_{L2}(Z)}{M_p} L_e^{\dagger}
(L_{\mu} \chi_1 +L_{\tau}  \chi_2 )+h.c. ,
\end{equation}
where $\al_{L \pm }(Z)$ and $\beta_{L i}(Z)$ are dimensionless 
generic functions of moduli fields $Z$, 
appear for the mixing between $L_e$ and $L_{\mu,\tau}$.
In addition, we have the following corrections 
\begin{equation}
\Delta K_L^{(\tau \mu)} = \frac{ \alpha_{L-}(Z)}{M_p^2}
(L^{\dagger}_{\mu}L_{\tau}-L^\dagger_{\tau}L_{\mu})\sigma_{--}
+  \frac{ \alpha_{L+}(Z)}{M_p^2}(L^{\dagger}_{\mu}L_{\tau}
+L^\dagger_{\tau}L_{\mu})\sigma_{+-} + \cdots ,
\label{K-corr}
\end{equation}
for the mixing between $L_\mu$ and $L_\tau$, where 
$\al_{L \pm }(Z)$ and $\beta_{L i}(Z)$ are generic dimensionless
functions of moduli fields $Z$. 
Here,  $\sigma_{--}$ and $\sigma_{+-}$ 
denote
\beq
\sigma_{--} = 
\chi_1^{\dagger} \chi_2 -
\chi_2^{\dagger} \chi_1, \qquad 
\sigma_{+-} = 
\chi_1^{\dagger} \chi_2 + 
\chi_2^{\dagger} \chi_1,
\eeq
and the ellipsis in (\ref{K-corr}) denotes terms including 
other bi-linear combinations of $\chi_1$, $\chi_2$ and their 
conjugates.  The contribution of the $D_4$ singlets $\chi$, $\chi_{-+}$ 
are suppressed by 
$Z_2$ symmetry and do not contribute to the leading order
of  off diagonal matrix elements. 
Also, the K\"ahler potential for the right handed 
charged leptons $R_{e,\mu,\tau}$ has similar corrections.
Including these corrections, scalar masses squared are 
estimated in the $D_4$ flavor basis as 
\beq
{m}_{L}^2 =
\begin{pmatrix} 
m_{L1}^2
& {\cal O}(\al m_{3/2}^2) & 
{\cal O}(\al m_{3/2}^2)  \\ 
{\cal O}(\al m_{3/2}^2)  & m_{L2}^2 
& {\cal O}(\al^2 m_{3/2}^2)  \\ 
{\cal O}(\al m_{3/2}^2)  & {\cal O}(\al^2 m_{3/2}^2)  &  
m_{L2}^2 
\end{pmatrix},
\eeq
\beq
{m}_{R}^2 =
\begin{pmatrix} 
m_{R1}^2
& {\cal O}(\al m_{3/2}^2) & 
{\cal O}(\al m_{3/2}^2)  \\ 
{\cal O}(\al m_{3/2}^2)  & m_{R2}^2
 &  {\cal O}(\al^2 m_{3/2}^2) \\ 
{\cal O}(\al m_{3/2}^2)  & {\cal O}(\al^2 m_{3/2}^2) &  
m_{R2}^2 
\end{pmatrix} ,
\eeq
\noindent
where corrections of ${\cal O}(\alpha^2 m^2_{3/2},
\alpha_a^2 m^2_{3/2},\alpha_b^2 m^2_{3/2},\alpha_a\alpha_b m^2_{3/2})$
are omitted in the diagonal elements.
Even though we include these corrections, we have 
almost the same constraint due to FCNC as 
the previous estimation (\ref{constraint-12}).

Also we examine the mass matrix between the left-handed and
the right-handed sleptons, which is generated by the so-called
A-terms.
The A-terms are trilinear couplings of two sleptons and one Higgs field, 
and are obtained as \cite{Kaplunovsky:1993rd}
\begin{eqnarray}
h_{IJ} {D}_I {R}_J H^d &= &  \tilde{h}_{IJ}{D}_I {R}_J H^d - 
(\tilde{y}_\ell)_{LJ} {D}_I {R}_J H^d F^{\Phi_k} K^{L\ov{L}}
\pa_{\Phi_k} K_{\ov{L}I} \nonumber \\  
& & -
(\tilde{y}_\ell)_{IM} {D}_I {R}_J H^d F^{\Phi_k} K^{M\ov{M}} 
\pa_{\Phi_k} K_{\ov{M}J}  ,
\label{eq:A-term}
\end{eqnarray}
where $\tilde h_{IJ} = F^{\Phi_k} \pa_{\Phi_k} (\tilde{y}_\ell)_{IJ}$.
Note that effective Yukawa couplings $(\tilde{y}_\ell)_{IJ}$ include 
$\chi,~\chi_{-+},~\chi_{i}$ as Eq.~(\ref{eq:clepton}).
After electroweak symmetry breaking, these provide us 
with the left-right mixing mass squared $(m^2_{LR})_{IJ} = h_{IJ}v^d$.

Now, let us discuss the first term in Eq.~(\ref{eq:A-term}), $\tilde h_{IJ} $.
For simplicity, we assume that Yukawa couplings are independent of 
moduli fields $Z$.
For example, the (3,3) element, $\tilde h_{33} $ is obtained as 
\beq
\tilde h_{33} = y_\mu \frac{F^{\chi}}{M_p} 
+ y'_\mu \frac{F^{\chi_{-+}}}{M_p}.
\eeq
Then, we can estimate $\tilde h_{33}v^d={\cal O}(m_\tau m_{3/2})$.
Similarly, we obtain the (2,2) element as 
\beq
\tilde h_{22} = y_\mu \frac{F^{\chi}}{M_p} - 
y'_\mu \frac{F^{\chi_{-+}}}{M_p}.
\eeq
Thus, we evaluate $\tilde h_{22}v^d={\cal O}(m_\tau m_{3/2})$, 
because we estimate 
$F^{\chi}/\chi, F^{\chi_{-+}}/\chi_{-+}$ $= {\cal O}(m_{3/2})$, but 
$F^{\chi}/\chi$ and $F^{\chi_{-+}}/\chi_{-+}$ are, 
in general, different from each other.
That may cause a problem.
If $|h_{IJ}/\tilde y_{IJ}|$ is large compared with 
slepton masses, there would be a minimum, where 
charge is broken~\cite{CCBbound}.\footnote{
If a decay rate from the realistic minimum to such 
charge breaking minimum is sufficiently small compared 
with the age of the universe, 
that might not be a problem. }
In order to avoid this, we assume that the K\"ahler metric of 
$\chi$ and $\chi_{-+}$ are the same,\footnote{
The $D_4$ flavor structure can be realized in heterotic string 
models on factorizable orbifolds including $Z_4$~\cite{Kobayashi:2006wq}.
In those heterotic orbifold models, $D_4$ non-trivial singlets 
and trivial singlets appear in the same sector and have 
the same K\"ahler metric.
In such models, our assumption would be justified. } 
e.g. canonical, and the non-perturbative 
superpotential leading to SUSY breaking does not include 
$\chi$ or $\chi_{-+}$, i.e. $\lan \partial_\chi W \ran = 
\lan \partial_{\chi_{-+}} W \ran =0$.
In this case, we obtain 
\beq
F^{\chi}/\chi = F^{\chi_{-+}}/\chi_{-+} = -m_{3/2},
\eeq
and we estimate $\tilde h_{22}v^d={\cal O}(m_\mu m_{3/2})$.
Similarly, other elements of $\tilde h_{IJ}$ are estimated as 
\beq
\tilde h_{IJ}v^d =
 m_{3/2}
  \begin{pmatrix}
{\cal O}(m_{e})   & {\cal O}(m_{\mu}\theta^\ell_{12}) &  
{\cal O}(m_{\tau}\alpha)  \\ 
{\cal O}(m_{\mu}\theta^\ell_{12})   & {\cal O}(m_{\mu})  & 
{\cal O}(m_{\tau}\alpha^2)   \\
{\cal O}(m_{\tau}\alpha)   & {\cal O}(m_{\tau}\alpha^2) & 
{\cal O}(m_{\tau})     \\  
\end{pmatrix}. 
\label{A-term-1}
\eeq
The other terms in Eq.~(\ref{eq:A-term}) lead to the same order of 
A-terms as Eq.~(\ref{A-term-1}).
Thus, we obtain the left-right mixing slepton mass matrix 
in the $D_4$ flavor basis as 
\beq
m^2_{LR} \equiv h_{IJ}v^d = m_{3/2}
  \begin{pmatrix}
{\cal O}(m_{e})   & {\cal O}(m_{\mu}\theta^\ell_{12}) &  {\cal O}(m_{\tau}\alpha)  \\ 
{\cal O}(m_{\mu}\theta^\ell_{12})   & {\cal O}(m_{\mu})  & {\cal O}(m_{\tau}\alpha^2)   \\
{\cal O}(m_{\tau}\alpha)   & {\cal O}(m_{\tau}\alpha^2) & {\cal O}(m_{\tau})     \\  
\end{pmatrix}. 
\eeq
The pattern of this mass matrix in the super-CKM basis 
is the same as the above.
We define the mass insertion parameters for the left-right mixing as 
\beq
(\delta^l_{LR})_{ij} \equiv \frac{(m_{LR}^2)_{ij}^{(SCKM)}}{m_{SUSY}^2} ,
\eeq
where $(m_{LR}^2)^{(SCKM)}$ is the left-right mixing slepton mass
squared matrix in the super-CKM basis.
Only for the (1,2) element of $(\delta^l_{LR})_{ij}$,
 there is a strong constraint due to FCNC as 
$(\delta^l_{LR})_{12} \leq {\cal O}(10^{-6})$~\cite{FCNCbound}, 
when $m_{SUSY}={\cal O}(100)$ GeV.  
This constraint also 
requires $\theta^\ell_{12} \leq {\cal O}(10^{-3})$, which 
is the same as Eq.~(\ref{constraint-12}).

The soft SUSY breaking terms, which we have studied, are 
generated at a high energy scale such as the Planck scale 
or the GUT scale.
In the above discussion, we have neglected 
radiative corrections.
The gaugino contributions are dominant in 
radiative corrections to slepton masses, that is, 
slepton masses at the weak scale are obtained by 
ones at the GUT scale $M_X$ as 
\begin{eqnarray}
m_{L}^2(M_Z) &= & m_{L}^2(M_X) + 0.5 M_{\tilde W}^2 
+ 0.04M_{\tilde B}^2, 
\nonumber \\
m_{R}^2(M_Z) &= & m_{R}^2(M_X) + 0.2 M_{\tilde B}^2, 
\end{eqnarray}
where $M_{\tilde B}$ and $M_{\tilde W}$ are bino and wino masses, 
respectively.
The above estimation on FCNC constraints does not change 
drastically when these gaugino masses are 
comparable with slepton masses.
If these gaugino masses are quite large compared with 
slepton masses, FCNC constraints would be improved.

\section{Conclusion}

We have studied supersymmetric extension of the 
$D_4 \times Z_2$ flavor model of~\cite{Ishimori:2008gp}.
We have evaluated soft SUSY breaking terms about the slepton mass
terms. 
 It is remarked that the second and third families of slepton masses are
almost  degenerate. The difference is tiny as 
${\cal O}(\alpha^2 m^2_{3/2},
\alpha_a^2 m^2_{3/2},\alpha_b^2 m^2_{3/2},\alpha_a\alpha_b m^2_{3/2})$.
The (1,2) element $\theta^\ell_{12}$ of the diagonalizing matrix for  
the charged lepton mass matrix is important for the FCNC constraints, 
in particular $\mu \rightarrow e \gamma$ experiments, 
although it is not important to realize the neutrino oscillation experiments.
It is constrained as $\theta^\ell_{12} \leq {\cal O}(10^{-3})$  from 
the current bound of $BR(\mu \rightarrow e \gamma )$
and our model is marginal.
Thus, future improvement on the bound of 
$BR(\mu \rightarrow e \gamma)$,  e.g. by the MEG experiment 
\cite{Mori:2007zza} is quite important in our model.

In Appendix, we have also discussed supersymmetric extension of 
the Grimus-Lavoura  $D_4 \times Z_2$ flavor model, which leads 
to almost the same results for soft SUSY breaking terms.

Finally, we give a comment on realization of our SUSY model 
by string model building.
The $D_4$ flavor symmetry can appear from heterotic string models 
on factorizable orbifold models including the $Z_2$ orbifold 
like $Z_2 \times Z_N$ 
orbifolds \cite{Kobayashi:2004ya,Kobayashi:2006wq,Ko:2007dz}, 
unless one does not introduce Wilson lines, which break 
degeneracy of massless spectra.
Indeed, several semi-realistic models have been 
constructed \cite{Kobayashi:2004ya,Kobayashi:2004ud}, where 
three families correspond to $D_4$ trivial singlets and 
doublets.
Stringy realization of our flavor structure 
would be plausible from this viewpoint.
However, such heterotic orbifold models include only 
$D_4$ trivial singlets and 
doublets, but not $D_4$ non-trivial singlets as fundamental modes.
On the other hand, the $D_4$ non-trivial singlet $\chi_{-+}$ 
plays an important role in our model.
Such a mode could appear as a composite mode.
Alternatively, $D_4$ non-trivial singlets as well as 
trivial singlets and doublets can appear as fundamental modes in 
heterotic string models on factorizable orbifolds including 
the $Z_4$ orbifolds like $Z_4 \times Z_N$ orbifolds.
Thus, stringy realization on such orbifolds might be alternative 
possibility.

\vspace{1cm}
\noindent
{\bf Acknowledgement}

T.~K.\/ is supported in part by the
Grand-in-Aid for Scientific Research, No. 17540251 and 
the Grant-in-Aid for
the 21st Century COE ``The Center for Diversity and
Universality in Physics'' from the Ministry of Education, Culture,
Sports, Science and Technology of Japan.
The work of R.T. has been  supported by the Japan Society of Promotion of
Science. 
The work of M.T. has been  supported by the
Grant-in-Aid for Science Research
of the Ministry of Education, Science, and Culture of Japan
Nos. 17540243 and 19034002.

\appendix

\section{Appendix}

Here, we discuss the supersymmetric extension of the Grimus-Lavoura 
model~\cite{Grimus}. 
The Grimus-Lavoura non-SUSY model includes three electroweak 
Higgs fields.
Thus, we have to introduce three pairs of Higgs superfields 
for the up and down sectors, $H^{u,d}_i$ ($i=1,2,3$).
Also we introduce a $D_4$ doublet $(\chi_1, \chi_2)$.
Table 2 shows $D_4$ and $Z_2$ charges for all fields.
\begin{table}[htb]
\begin{center}
\begin{tabular}{|c|cccccc||ccc||c|}
\hline
              &$L_e$   & $L_I$ & $R_e$  & $R_I$
              & $N_e$ & $N_I$ &  $H^{u,d}_1$&$H^{u,d}_2$  &
$H^{u,d}_3$   &$(\chi_1,\chi_2)$   \\ \hline
$D_4$      &${\bf 1}_{++}$       & ${\bf 2}$                    & ${\bf
  1}_{++}$      &                ${\bf 2}$ &       ${\bf 1}_{++}$ & ${\bf 2}$
& ${\bf 1}_{++}$         &${\bf 1}_{++}$&${\bf 1}_{+-}$& ${\bf 2}$      \\
$Z_2$ &   +     &          +              &  $-$ &      +
&     $-$        &           $-$                & $-$ & + & + & + \\ \hline
\end{tabular}
\end{center}
\caption{$D_4$ and $Z_2$ charges for the supsymmetric Grimus-Lavoura
  model. $I$ corresponds to $\mu$ and $\tau$.
\label{charge}}
\end{table}

Following these assignments, the superpotential which leads 
the lepton mass matrices is found to be 
\beq
\begin{split}
W^{(3)}=&~ y_e L_e R_e H^d_1 +y_\mu (L_{\mu}R_{\mu}+D_{\tau}R_{\tau})
H^d_2 +y'_\mu (L_{\tau}R_{\tau}- L_{\mu}R_{\mu}) H^d_3 \\
  &+(y_1 L_e N_e +y_2 (L_{\mu}N_{\mu}+L_{\tau}N_{\tau})) H_1^u 
 + y_3 ( L_{\tau}N_{\tau}-L_{\mu}N_{\mu})H^u_3\\
  &+y_a N_e(N_{\mu}\chi_1+N_{\tau}\chi_2) 
  +M_1 N_eN_e+M_2(N_{\mu}N_{\mu}+N_{\tau}N_{\tau}),
\end{split}  
\eeq
up to 4-point couplings.
Furthermore, higher-order superpotential terms like,
\beq
W^{(4)}=y_{e\mu}(\chi_1 R_{\mu}+\chi_2 R_{\tau})L_e H_2^d + \cdots,
\eeq
 should be also considered, because $\lan \chi_i \ran$ must be large
 to lead the realistic neutrino mixing. 
Now the charged lepton mass matrix can be evaluated as follows 
\beq  
M_l
= \begin{pmatrix}
y_e v^d_1  &  (y_{e\mu}v^d_2-y'_{e\mu}v^d_3)\alpha &  
(y_{e\mu}v^d_2+y'_{e\mu}v^d_3)\alpha \\ 
y_{\mu e} v^d_1 \al & y_{\mu} v^d_2 -y'_{\mu} v^d_3  & {\cal O}(v^d_2 \alpha^2)     \\
y_{\mu e} v^d_1 \al & {\cal O}(v^d_3\alpha^2) & y_{\mu} v^d_2 +y'_{\mu} v^d_3     
\end{pmatrix},
\eeq
where $\lan H^d_i \ran =v^d_i$. If $\al \ll 1$ is allowed, eigenvalues of 
lepton masses can be found to be equal to the diagonal elements like
Eq.~(\ref{eq:clepton}). 
We need fine-tuning Yukawa couplings and VEVs such that 
$(y_{\mu} v^d_2 -y'_{\mu} v^d_3)/(y_{\mu} v^d_2 +y'_{\mu} v^d_3)=
{\cal O}(m_\mu/m_\tau)$.
In addition, we require $y_e v^d_1$ to be suppressed 
compared with $y_{\mu} v^d_2 +y'_{\mu} v^d_3$ to lead to 
the mass ratio $m_e/m_\tau$, i.e. 
$y_e v^d_1/(y_{\mu} v^d_2 +y'_{\mu} v^d_3)=
{\cal O}(m_e/m_\tau)$.
The (1,2) element of diagonalizing matrix $\theta^\ell_{12}$ 
is estimated as $\theta^\ell_{12}=
(y_{e\mu}v^d_2-y'_{e\mu}v^d_3)\alpha/m_\mu$ and 
it reduces to $\theta^\ell_{12} \sim y_{e \mu}\alpha m_\tau/m_\mu$ 
for $y_{e \mu} \sim y'_{e \mu}$ and $v^d_2 \sim v^d_3$.
Thus, this is also the same as Eqs.~(\ref{theta-12-1}) and 
(\ref{theta-12-2}).

On the other hand, the Dirac mass matrix, $M_D$, and 
the Majorana mass matrix, $M_R$, 
in the neutrino sector are written as 
\begin{eqnarray}
M_D
&=& 
\left(
  \begin{array}{ccc}
                  y_{1} v^u_1    & (y_{12}v^u_2-y'_{12}v^u_3)\alpha  & (y_{12}v^u_2+y'_{12}v^u_3)\alpha  \\ 
                    y_{21} v^u_1 \alpha    & y_2v^u_1-y_3v^u_3  & {\cal O}( v_3^u \alpha^2)    \\
                    y_{21} v^u_1 \alpha     &{\cal O}( v^u_3\alpha^2 ) & y_2v^u_1+y_3v^u_3   \\
  \end{array} \right),
\nonumber\\
M_R
&=& 
\left(
  \begin{array}{ccc}
M_1   &  y_a M_p \alpha &  y_a M_p\alpha  \\ 
           y_a   M_p \alpha    & M_2  & y_{b}M_p \alpha^2    \\
y_a M_p \alpha   & y_{b}M_p \alpha^2 & M_2     \\  
\end{array} \right),
\end{eqnarray}
where $\lan H^u_i \ran =v^u_i$. 
The above pattern is quite similar to Eq.~(\ref{MR-MD}), and in particular, 
the form of $M_R$ is the same as Eq.~(\ref{MR-MD}).
Thus, similar values of parameters lead to realistic results 
(\ref{best-fit}), 
i.e.,  $M_1$, $M_2 = {\cal O}(10^{15})$GeV and  $\al \sim M_2/M_p$.
Thus the favorable region of $\alpha$ is of 
${\cal O}(10^{-4})-{\cal O}(10^{-2})$.

In this model, the soft SUSY breaking terms are also restricted by
$D_4 \times Z_2$ symmetry and  expected not to be different from the
estimation in Section 3. 
In fact, the scalar mass terms are given by  
\beq
{m}_{L}^2 =
\begin{pmatrix} 
m_{L1}^2+{\cal O}(\al^2 m_{3/2}^2)  & {\cal O}(\al m_{3/2}^2) & 
{\cal O}(\al m_{3/2}^2)  \\ 
{\cal O}(\al m_{3/2}^2)  & m_{L2}^2 +{\cal O}(\al^2 m_{3/2}^2)     & 
{\cal O}(\al^2 m_{3/2}^2)  \\ 
{\cal O}(\al m_{3/2}^2)  & {\cal O}O\al^2 m_{3/2}^2)  &  
m_{L2}^2 + {\cal O}(\al^2 m_{3/2}^2)
\end{pmatrix},
\eeq
\beq
{m}_{R}^2 =
\begin{pmatrix} 
m_{R1}^2+{\cal O}(\al^2 m_{3/2}^2)  & 0 & 0  \\ 
0  & m_{R2}^2+{\cal O}(\al^2 m_{3/2}^2)     &  {\cal O}(\al^2 m_{3/2}^2) \\ 
0  & {\cal O}(\al^2 m_{3/2}^2) &  m_{R2}^2 + {\cal O}(\al^2 m_{3/2}^2)
\end{pmatrix},
\eeq
in the $D_4$ flavor basis.
Thus, we obtain the same constraint on the mass insertion 
parameters due to FCNC as (\ref{constraint-12}), i.e. 
$\theta^\ell_{12} \leq {\cal O}(10^{-3})$.

Similarly, the A-terms can also be evaluated.
This model would have the same problem about the 
(2,2) element of A-terms as the model in Section 3, 
that is, the (2,2)-element would be of 
${\cal O}(m_{\tau}m_{3/2})$ without tuning 
about the coefficient. 
However, when the K\"ahler metric of $H^d_2$ and $H^d_3$ are the same, 
the (2,2) element becomes of ${\cal O}(m_{\mu}m_{3/2})$.
Then, the left-right mixing slepton mass matrix could be estimated as  
\beq
m_{LR}^2 =m_{3/2}
\begin{pmatrix} 
{\cal O}(m_e) & {\cal O}( \theta^\ell_{12} m_\tau ) & 
{\cal O}(\al m_\tau  ) \\ 
{\cal O}(\al m_e ) & {\cal O}(m_\mu )    & 
{\cal O}(\al^2 m_\tau ) \\ 
{\cal O}(\al m_e ) & {\cal O}(\al^2 m_\tau ) & 
{\cal O}( m_\tau ) 
\end{pmatrix},
\eeq
in the $D_4$ flavor basis, and its form is almost the same in 
the super-CKM basis.
Thus, the mass insertion parameter is estimated as 
$ (\delta_{LR}^l)_{12} = {\cal O}(\theta^\ell_{12} m_{\mu}/ m_{3/2})$, and 
we have the same constraint as one in Section 3, 
i.e. $\theta^\ell_{12} \leq {\cal O}(10^{-3})$.

As a result, soft SUSY breaking terms, which are predicted 
in the supersymmetric Grimus-Lavoura model, are 
almost the same as those obtained in Section 3.
The difference is the number of Higgs pairs, 
that is, the supersymmetric Grimus-Lavoura model has 
three pairs of Higgs supermultiplets, while 
the model in Section 2 has only one pair and it becomes 
the MSSM at low energy.
The former may violate the gauge coupling unification 
unless one introduces extra colored supermultiplets.

\end{document}